\documentclass[12pt]{article}
\usepackage{amsmath,amssymb,bm,epsfig}
\renewcommand{\thefootnote}{\fnsymbol{footnote}}

\begin{document}

\title{
\begin{flushright}
\begin{minipage}{0.2\linewidth}
\normalsize
EPHOU-16-016 \\
WU-HEP-16-017 \\*[50pt]
\end{minipage}
\end{flushright}
{\Large \bf 
Constraints on small-field axion inflation
\\*[20pt]}}

\author{Tatsuo~Kobayashi,$^{1,}$\footnote{
E-mail address: kobayashi@particle.sci.hokudai.ac.jp}\ \ 
Akane~Oikawa,$^{2,}$\footnote{
E-mail address: a.oikawa@aoni.waseda.jp}\ \ 
Naoya Omoto,$^{1,}$\footnote{
E-mail address: omoto@particle.sci.hokudai.ac.jp}\\ 
Hajime~Otsuka,$^{2,}$\footnote{
E-mail address: h.otsuka@aoni.waseda.jp}
\ and \
Ikumi~Saga$^{1,}$\footnote{
E-mail address: i-saga@particle.sci.hokudai.ac.jp}
\\*[20pt]
$^1${\it \normalsize 
Department of Physics, Hokkaido University, Sapporo 060-0810, Japan}  \\
$^2${\it \normalsize 
Department of Physics, Waseda University, 
Tokyo 169-8555, Japan} \\*[50pt]}

\date{
\centerline{\small \bf Abstract}
\begin{minipage}{0.9\linewidth}
\medskip 
\medskip 
\small
We study general class of small-field axion inflations which are 
the mixture of polynomial and sinusoidal functions suggested by the 
natural and axion monodromy inflations. 
The axion decay constants leading to the successful axion inflations 
are severely constrained in order not to spoil the Big-Bang nucleosynthesis 
and overproduce the isocurvature perturbation originating 
from the QCD axion. 
We in turn find that the cosmologically favorable axion decay constants are typically of order 
the grand unification scale or the string scale which is consistent 
with the prediction of closed string axions. 
\end{minipage}
}

\begin{titlepage}
\maketitle
\end{titlepage}

\renewcommand{\thefootnote}{\arabic{footnote}}
\setcounter{footnote}{0}

\section{Introduction}
An axion is an attractive candidate for the inflation scenario in addition to  other 
phenomenologically favorable scenarios such as a solution of the 
strong CP problem and the candidate of dark matter in our Universe. 
Furthermore, string theory predicts a lot of axion particles in the low-energy effective theory 
through the compactification of extra-dimensional space. 
When the axion is associated with the higher-dimensional form fields, 
the form of the axion potential is protected by the higher-dimensional gauge symmetries 
apart from other matter fields. 
The axion potential is then generated by the spontaneous or explicit breaking of axionic shift symmetry 
originating from the higher-dimensional one. 
The higher-dimensional operators in the axion potential are still controlled, 
and the cosmological observables induced by the axion inflation are predictable. 

To construct the inflationary favorable axion potential, the axion decay 
constant is required to be enough large to obtain the flat direction in the axion potential, 
in particular, the trans-Planckian decay constant for the natural inflation~\cite{Freese:1990rb}. 
However, in string theory, the decay constant of a closed string axion is typically around the 
string scale or grand unification scale $10^{16}\,{\rm GeV}$~\cite{Choi:1985je,Banks:2003sx,Svrcek:2006yi}. 
When the axion decay constant is of order the Planck scale, 
the axion inflation generically predicts ${\cal O}(1)$ tensor-to-scalar ratio 
as can be seen in the Lyth bound~\cite{Lyth:1996im}, 
which argues that the tensor-to-scalar ratio $r$ is closely related to the inflaton field range, 
$\Delta \phi$, during the inflation. 
Under the assumption that a variation of $r$ is negligible over the period $\Delta \phi$, 
the approximate relation is obtained as~\cite{Lyth:1996im},
\begin{eqnarray}
\frac{\Delta \phi}{M_{\rm Pl}} \simeq {\mathcal O}(1) \times \left(\frac{r}{0.01}\right)^{1/2}.
\label{eq:Lyth_bound}
\end{eqnarray}
This indicates that if $\Delta\phi < M_{\rm Pl}$, $r\lesssim 0.01$ is obtained and we call 
this class of inflation model the small-field inflation throughout this paper. 
Although the large-field axion inflations ($r\gtrsim 0.01$) are consistent with the recent Planck 
data~\cite{Planck:2013jfk,Ade:2015lrj}, 
the weak gravity conjecture~\cite{ArkaniHamed:2006dz} suggests that the higher-order instanton 
effects give a sizable effect for the axion potential with a trans-Planckian axion decay constant and 
these would generically violate the slow-roll axion inflation. 

In this respect, we consider the axion inflations with the decay constant below the Planck scale 
or string scale, which are favorable from the aspects of the weak gravity conjecture. 
The flat direction required in the inflation can be realized by 
choosing the proper parameters in the axion potential. 
Since the obtained inflaton potential is categorized into the class of small-field axion inflation, 
it predicts the small amount of gravitational waves and low inflation scale 
in comparison with the prediction of large-field axion inflation. 
Such a low-scale inflation is also influential to the isocurvature perturbation originating 
from the QCD axion. 
When all the dark matter is dominated by the QCD axion, 
the current Planck result constrains the Hubble scale during the inflation 
$H_{\rm inf}$~\cite{Planck:2013jfk}, 
\begin{align}
H_{\rm inf} <0.87 \times 10^7{\rm GeV}\left(\frac{f_{\rm QCD}}{10^{11}\,{\rm GeV}}\right)^{0.408},
\label{eq:iso}
\end{align}
where $f_{\rm QCD}$ is the decay constant of the QCD axion. 
It is then possible to avoid the isocurvature constraint by the low-scale inflation, 
although the upper bound of $f_{\rm QCD}$ depends on the initial misalignment angle of the axion 
and dilution mechanism after the inflation~\cite{Kawasaki:2004rx,Hattori:2015xla,Akita:2016usy}.  
 
Recently, some of the authors  conjectured that, in a certain class of small-field axion inflation 
derived from type IIB superstring theory~\cite{Kobayashi:2015aaa},\footnote{
The model in Ref.~\cite{Kobayashi:2015aaa} can lead to both small-field and large-field inflations.} 
the tensor-to-scalar ratio $r$ correlates with the axion decay constant $f$ as 
follows~\cite{Kadota:2016jlw},
\begin{align}
r \sim 10^{-6} f^{2q},
\nonumber
\end{align}
where the fractional number $q$ depends on the model. 
In the example of Refs.~\cite{Kobayashi:2015aaa,Kadota:2016jlw}, we obtain $q=2$.
This behavior originates from sinusoidal functions in the axion inflation potential.
The above relation could also predict the magnitude of the inflation potential 
and the inflaton mass by $f$.
In general, superstring theory leads to the axion potential with one or more sinusoidal terms 
induced by several non-perturbative terms.
Thus, it is important to extend the previous analysis to other axion inflation scenarios. 
In this paper, we further study such dependence of the axion decay constant for 
not only cosmological observables, but also the reheating temperature and 
dark matter abundance for the general class of small-field axion inflations, which 
are the mixture of polynomial and sinusoidal functions suggested in the axion monodromy inflation~\cite{Silverstein:2008sg,Kobayashi:2014ooa,Higaki:2014sja} 
and general form of sinusoidal functions suggested in the natural and multi-natural inflations~\cite{Freese:1990rb,Choi:2014rja,Czerny:2014wza,Czerny:2014xja}.\footnote{The scalar potential including modular functions 
in superstring theory can effectively lead to such a multi-natural inflation \cite{Abe:2014xja}.} 
We constrain the axion decay constant realizing the small-field axion inflations 
by the isocurvature perturbation originating from the 
QCD axion, successful Big-Bang nucleosynthesis (BBN) and dark matter abundance. 
As will be shown, it is quite interesting that the allowed range of the axion decay constant 
corresponds to the typical decay constant region realized in  superstring theory, 
when our axion is the closed string axion~\cite{Choi:1985je,Banks:2003sx,Svrcek:2006yi}.

In the remainder of this paper, we first discuss the conditions leading to the 
general class of small-field axion inflations and analytical form of cosmological 
observables as a function of the decay constant in Sec.~\ref{sec:2}. 
In Sec.~\ref{sec:3}, we derive the constraints for the axion decay constants from the 
reheating process and dark matter abundance. We summarize our conclusion in Sec.~\ref{sec:con}.

\section{Small-field axion inflation}
\label{sec:2}
In this section, we consider the small-field axion inflations with an emphasis on 
the multi-natural inflation in Sec.~\ref{subsec:multi} and axion monodromy inflation 
with sinusoidal functions in Sec.~\ref{subsec:mono}. 
The current Planck results constrain the power spectrum of 
curvature perturbation $P_\xi$, its spectral index $n_s$, and 
tensor-to-scalar ratio $r$~\cite{Planck:2013jfk,Ade:2015lrj}, written in terms of 
the slow-roll parameters $\epsilon = \frac{1}{2} \left(\frac{V_{\phi}}{V}\right)^2$ 
and $\eta = \frac{V_{\phi \phi}}{V}$ with $V_{\phi}=\partial_\phi V$ ($V_{\phi\phi}=\partial_\phi\partial_\phi V$) 
being the first (second) derivative of the inflaton potential $V(\phi)$\footnote{Here and in what follows, 
we use the reduced Planck unit $M_{\rm Pl}=2.4\times 10^{18}\,{\rm GeV}=1$ unless otherwise specified.}, 
\begin{align}
P_{\xi} &=\frac{V}{24 \pi^2 \epsilon}= 2.20\pm0.10 \times 10^{-9},
\nonumber\\
n_{s} &= 1+2\eta-6\epsilon= 0.9655 \pm 0.0062,
\nonumber\\
r &=16\epsilon< 0.12.
\label{eq:obs}
\end{align}
We derive the constraints for the parameters in the axion potentials leading to the 
successful small-field axion inflation. 

\subsection{Multi-natural inflation}
\label{subsec:multi}
First of all, we proceed to study the extended natural inflation, so-called multi-natural inflation~\cite{Czerny:2014wza,Czerny:2014xja}, 
in which the general form of inflaton potential is given by
\begin{align}
V(\phi) &= \sum^{M}_{m=1} A_m \cos \left(\frac{\phi}{f_m}+\theta_m \right) + V_0. 
\label{eq:multi_inf}
\end{align}
Here, $\phi$ is a canonically normalized axion with the decay constants $f_m$ 
with $m=1,2,\cdots,M$; 
$\theta_m$ denotes the phase of sinusoidal functions; 
$A_m$ are the real positive constants; 
and $V_0$ is the real constant to achieve the tiny cosmological constant. 
$M$ depends on the number of hidden gauge sectors which non-perturbatively 
generate the potential of the axion inflaton.

Let us demonstrate the small-field axion inflation by the small axion decay constants $f_m$ 
with $m=1,2,\cdots, M$, 
where a sufficiently large number of {\it e}-folding is achieved under the flat direction in the axion potential. 
To achieve such a situation, the first derivative of potential in Eq.~(\ref{eq:multi_inf}),  
\begin{equation}
V_{\phi} =-\sum_{m=1}^M\frac{A_m}{f_m}  \sin\left(\frac{\phi}{f_m}+\theta_m\right),
\label{eq:first_multi_inf}
\end{equation}
is required to be smaller than its potential during the inflation, i.e., $|V_{\phi}| \ll |V|$. 
It can be realized with the region satisfying 
\begin{align}
&\sin(\phi/f_{m}+\theta_m )\sim \cos(\phi/f_{m}+\theta_m ) \sim {\cal O}(1),
\label{eq:assum1}
\end{align} 
with these proper signs and the correlated parameters in the scalar potential,
\begin{equation}
\frac{A_m}{f_m} \sim \frac{A_n}{f_n},
\label{eq:assum2}
\end{equation}
for any $m,n=1,2,\cdots,M$. 

Since the slow-roll inflation is realized under $|V_{\phi}| \simeq 0$ and $|V_{\phi \phi}| \simeq 0$ 
during the inflation, the second derivative of the potential can be estimated by employing 
the inflaton variation $\Delta \phi$, 
\begin{equation}
V_{\phi \phi} 
\sim V_{\phi \phi \phi} \Delta \phi 
\sim \left(
-\sum_{m=1}^M\frac{A_m}{f_m^3} \sin\left(\frac{\phi}{f_m}+\theta_m\right) \right)\Delta \phi.
\end{equation}
Here, we assume that all $f_{m}^{-3}$ can be dominated in the third derivative $V_{\phi \phi \phi}$.
With the help of Eqs.~(\ref{eq:assum1}) and~(\ref{eq:assum2}), the slow-roll parameter 
is obtained as
\begin{align}
\eta \sim 
\frac{\sum_{m=1}^M \frac{A_m}{f_m^3}}{\sum_{n=1}^M A_n} \Delta \phi  
\sim \left(\frac{\sum_m\frac{1}{f_m^2}}{\sum_{n} f_n} \right)\Delta \phi,
\end{align}
where $V\sim \sum_m A_m$ is employed. 
Since we concentrate on the parameter space leading to the small-field inflation, 
the slow-roll parameter $\epsilon$ is 
expected to be much smaller than unity. 
It is confirmed later by checking the value of the tensor-to-scalar ratio $r = 16 \epsilon$. 
Thus, slow-roll parameter $|\eta|$ is chosen as $10^{-2}$ to reproduce 
the observed spectral index $n_s\simeq 0.96$ reported by Planck. 

By fixing $|\eta| \simeq 10^{-2}$, the tensor-to-scalar ratio is estimated by using the 
Lyth bound Eq.~(\ref{eq:Lyth_bound}), 
\begin{align}
r \sim  10^{-2} \times (\Delta \phi)^2  
\sim  10^{-6} \times  \left(\frac{\sum_m\frac{1}{f_m^2}}{\sum_{n} f_n} \right)^{-2}
\times \left( \frac{\eta}{0.01}\right)^2.
\label{eq:r_multi}
\end{align}

Furthermore, we can estimate the energy scale of the scalar potential during the inflation $V_{\rm \inf}$ as a 
function of the axion decay constants from Eqs.~(\ref{eq:obs}) and~(\ref{eq:r_multi}),
\begin{align}
V_{\rm \inf}^{1/4} &\sim  4 \times 10^{-4} \times 
\left(\frac{\sum_{m} f_m}{\sum_n\frac{1}{f_n^2}} \right)^{1/2},
\label{eq:Vinf_multi}
\end{align}
and consequently the Hubble parameter $H_{\rm inf} = (V_{\rm inf}/3)^{1/2}$ becomes
\begin{align}
H_{\rm inf} \sim 10^{-7} \times \left(\frac{\sum_{m} f_m}{\sum_n\frac{1}{f_n^2}} \right).
\label{eq:H_multi}
\end{align}

Finally, we estimate the inflaton mass $m_{\phi}^2$ as a function of the decay constant.
For small $f_m$, the dominant term of the second derivative, $V_{\phi \phi}$, at $\phi=0$
is evaluated by using Eq.~(\ref{eq:assum1}), $V_{\phi \phi} \sim  \sum_m \frac{A_m}{f_m^2}$, 
and hereafter the inflaton mass is estimated as 
\begin{align}
m_{\phi}^2 =V_{\phi\phi} 
&\sim  \sum_m \frac{A_m}{f_m^2}
\sim  \frac{\sum_m \frac{A_m}{f_m^2}}{\sum_n A_n}V_{\rm inf} 
\sim  \left(\frac{\sum_m \frac{1}{f_m}}{\sum_n f_n}\right)V_{\rm inf} 
\sim  3\times 10^{-14} \left(\frac{\sum_m \frac{1}{f_m}}{\sum_n f_n}\right)\left(\frac{\sum_{n} f_n}{\sum_m\frac{1}{f_m^2}} \right)^2.
\label{eq:m_multi}
\end{align}

Let us summarize the result for two nonvanishing sinusoidal functions in Eq.~(\ref{eq:multi_inf}).  
For $f_{1} \sim f_{2} \sim f \ll 1$, the obtained physical quantities have the following decay constant 
dependence:
\begin{eqnarray}
r \sim  10^{-6}  \times f^6\ , \ \ \ \ 
V_{\rm \inf}^{1/4} \sim  4 \times 10^{-4}   \times f^{3/2}, \nonumber \\
H_{\rm inf} \sim 10^{-7} \times f^3\ , \ \ \ \ \ 
m_{\phi}^2 \sim 3 \times 10^{-14}  \times{f^4}.
\label{eq:typical parameter values of type 1}
\end{eqnarray}
For another case $f_{1} \gg f_{2} $, they are written as
\begin{eqnarray}
r \sim  10^{-6}  \times (f_1f_2^2)^2\ , \ \ \ \ 
V_{\rm \inf}^{1/4} \sim  4 \times 10^{-4}   \times (f_1f_2^2)^{1/2}, \nonumber \\
H_{\rm inf} \sim 10^{-7} \times (f_1f_2^2)\ , \ \ \ \ \ 
m_{\phi}^2 \sim 3 \times 10^{-14}  \times (f_1f_2^3).
\label{eq:typical parameter values of type 1}
\end{eqnarray}

\begin{figure}[htbp]
  \begin{center}
    \begin{tabular}{c}

      \begin{minipage}{0.5\hsize}
        \begin{center}
          \includegraphics[clip, width=7.0cm]{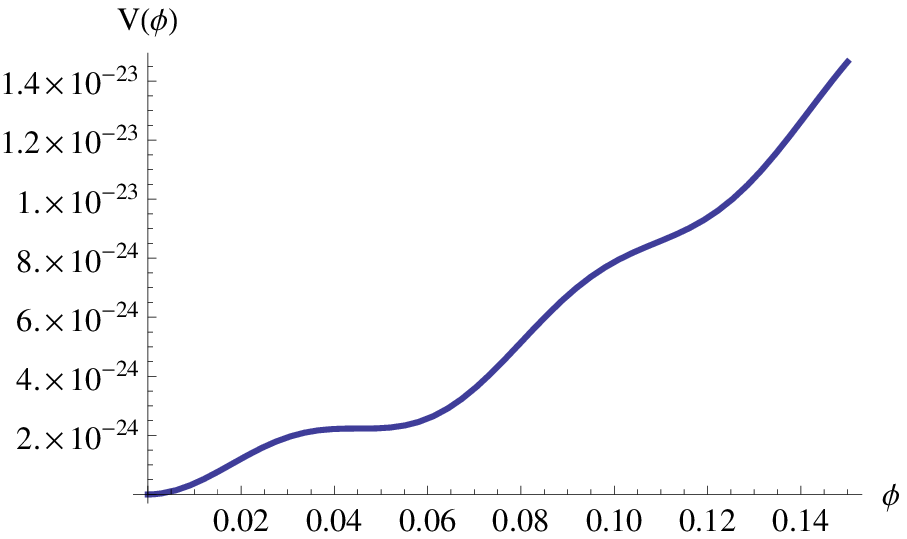}
          \hspace{1.6cm} 
        \end{center}
      \end{minipage}

      \begin{minipage}{0.5\hsize}
        \begin{center}
          \includegraphics[clip, width=7.0cm]{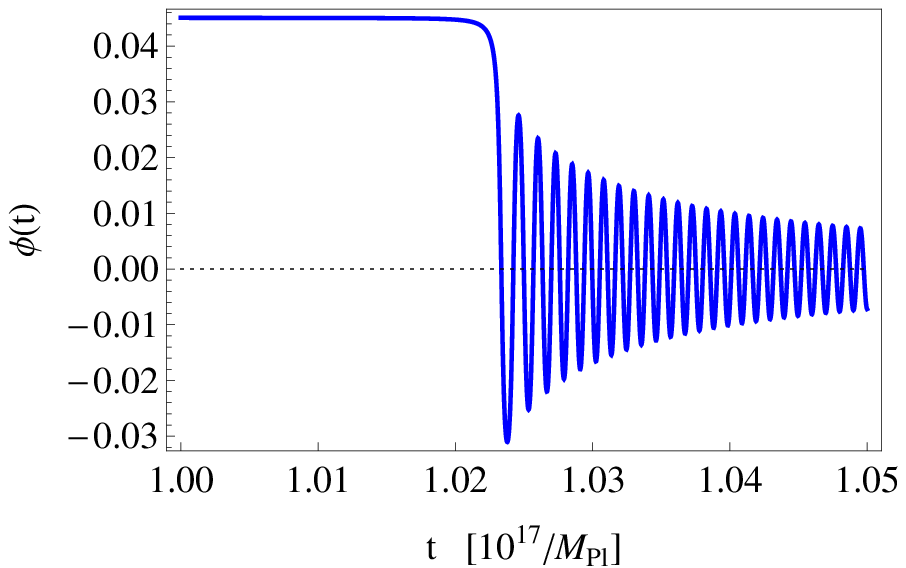}
          \hspace{1.6cm} 
        \end{center}
      \end{minipage}

    \end{tabular}
    \caption{In the left panel, the inflaton potential is drawn by setting the 
    parameters as $A_1/A_2 = 22.474579785926$ and $A_2=6.47\times 10^{-25}$, whereas the right panel 
    shows the trajectory of the inflaton as a function of cosmic time $t$ for 
    the initial value of the inflaton, $\phi_{\rm ini}= 0.04508690783$.}
    \label{fig:1}
  \end{center}
\end{figure}

Following this line of thought, we show the numerical analysis for specific axion potentials. 
For the illustrative purposes, we consider the axion potential with two sinusoidal 
functions\footnote{For details of bumpy natural inflation with the same scalar potential, see, 
Ref.~\cite{Parameswaran:2016qqq}.},
\begin{eqnarray}
V(\phi) &=& A_1 \left( 1 - \cos\left(\frac{\phi}{f_1}\right) \right)+ A_2 \left( 1 - \cos\left(\frac{\phi}{f_2}\right) \right),
\label{eq:multi_simplify}
\end{eqnarray}
which is achieved under $\theta_{1}=\theta_{2}=-\pi$ and $V_0\simeq A_1+A_2$ in Eq.~(\ref{eq:multi_inf}). 
For an illustrating example, we set the decay constants,  
$f_1=0.1$ and $f_2=0.01$. 
Fig.~\ref{fig:1} shows the inflaton potential and the trajectory of inflaton as a function of cosmic time $t$, 
where the parameters are set as $A_1/A_2 = 22.474579785926$ and $A_2=6.47\times 10^{-25}$. 
By solving the equation of motion for the inflaton field, we numerically obtain the 
cosmological observables as shown in Tab.~\ref{tab:1}. 
It is found that the analytical forms of physical quantities derived in Eqs.~(\ref{eq:r_multi})-(\ref{eq:m_multi}),
\begin{eqnarray}
r &\sim& 10^{-6}  \times \left(\frac{f_1 +f_2}{\frac{1}{f_1^2}+\frac{1}{f_2^2}} \right)^{2} 
   \sim  10^{-6}  \times (f_1 f_2^2)^2
   \sim 10^{-16}, \nonumber \\
V_{\rm \inf}^{1/4} &\sim&  4 \times 10^{-4}  \times \left(\frac{f_1 +f_2}{\frac{1}{f_1^2}+\frac{1}{f_2^2}} \right)^{1/2}
   \sim  4 \times 10^{-4} \times (f_1 f_2^2)^{1/2} 
   \sim 10^{-6}, \nonumber \\
H_{\rm inf} &\sim&  10^{-7} \times \left(\frac{f_1 +f_2}{\frac{1}{f_1^2}+\frac{1}{f_2^2}} \right) 
   \sim 10^{-7} \times (f_1 f_2^2) 
   \sim 10^{-12}, \nonumber \\
m_{\phi}^2 &\sim& 3 \times 10^{-14} \times \left( \frac{1}{f_1} + \frac{1}{f_2}\right)\left(\frac{f_1 +f_2}{\frac{1}{f_1^2}+\frac{1}{f_2^2}} \right)^{2}
   \sim 3 \times 10^{-14} \times \frac{1}{f_2}(f_1 f_2^2)^2
   \sim 3 \times 10^{-22},
\end{eqnarray}
are consistent with our obtained numerical results in Tab.~\ref{tab:1}.

\begin{table}[htb] 
 \begin{center}
  \begin{tabular}{|c|c|c|c|c|c|c|} \hline
     $N$ & $n_s$ & $r $ & 
     $m_{\phi}^2$ & $H_{\rm inf}$ &  $V_{\rm inf}^{1/4}$ & $\frac{dn_s}{d \ln k}$ \\ \hline\hline 
   60.0 & $0.9665$ & $6.85\times 10^{-17}$
   & $7.92\times 10^{-21}$& $8.62\times 10^{-13}$& $1.22\times 10^{-6}$ & $-2.52 \times 10^{-3}$ \\ 
  50.0 & $0.9665$ & $1.61\times 10^{-16}$ 
  & $1.86\times 10^{-20}$ & $1.32\times 10^{-12}$ & $1.51\times 10^{-6}$ & $-1.65 \times 10^{-3}$ \\ \hline
  \end{tabular}
      \caption{The cosmological observables such as spectral index $n_s$, its running $dn_s/d\ln k$, tensor-to-scalar ratio $r$, 
      Hubble scale $H_{\rm inf}$, scalar potential $V_{\rm inf}^{1/4}$ at the pivot scale and 
      the inflaton mass $m_{\phi}^2$ at the vacuum. 
      The parameters are set as $A_1/A_2 = 22.474579785926$ and $A_2=6.47\times 10^{-25}$ 
      for the {\it e}-folding number $N=60$, whereas those are set as $A_1/A_2 = 22.474579787160$ and 
      $A_2=6.47\times 10^{-25}$ 
      for the {\it e}-folding number $N=50$. The initial value of the inflaton field is also set 
      as $\phi_{{\rm ini}}=0.04508690783$ in both cases.}
    \label{tab:1}
  \end{center}
\end{table}

Tab.~\ref{tab:1} also shows numerical results of the running, $dn_s/d\ln k$, 
which are large and negative as in the case of Ref.~\cite{Parameswaran:2016qqq}.
These values can be estimated roughly as follows.
The slow-roll parameter, $\xi$, can be written
\begin{equation}
\xi = \frac{V_\phi V_{\phi \phi \phi}}{V^2} = \left(\frac{r}{8} \right)^{1/2}  \frac{ V_{\phi \phi \phi}}{V}.
\end{equation}
Then, using Eq.~(\ref{eq:r_multi}) and 
\begin{equation}
\frac{ V_{\phi \phi \phi}}{V} \sim \left(\frac{\sum_m\frac{1}{f_m^2}}{\sum_{n} f_n} \right),
\end{equation}
we can estimate $\xi ={\cal O}(10^{-3})$ for $\eta \sim 0.01$, and this value of $\xi$ is independent of decay constants.
In this model, the running is obtained as $dn_s/d\ln k \approx -2 \xi$ and other terms are sub-dominant.
Thus, it is found that $dn_s/d\ln k ={\cal O}(10^{-3})$.
Similarly, the running of running is obtained as $d^2n_s/d\ln k^2 \approx 2 \eta \xi$ in this model and other terms are sub-dominant.
Then, we find that $d^2n_s/d\ln k^2 ={\cal O}(10^{-5})$, which is also independent of decay constants.

Similarly, the potential (\ref{eq:multi_simplify}) with other values of 
$f_1$ and $f_2$ leads to results consistent with Eq.~(\ref{eq:typical parameter values of type 1}).

\subsection{Axion monodromy inflation with sinusoidal functions}
\label{subsec:mono}
We next discuss the axion monodromy inflation with sinusoidal functions 
in which the general form of the axion potential is yielded by
\begin{align}
V(\phi) &= A_1\phi^p + \sum^{M}_{i=2} A_i \cos \left(\frac{\phi}{f_i}+\theta_i \right) + V_0.
\label{eq:mono_inf}
\end{align}
Here, $\phi$ is a canonically normalized axion with the decay constants $f_i$; 
$\theta_i$ denotes the phase of the sinusoidal functions; 
$A_i$ are the real positive constants; 
and $V_0$ is the real constant to achieve the tiny cosmological constant. 
$p$ can be taken as fractional numbers or positive integers such as 
$p=1$~\cite{McAllister:2008hb}, $p=2/3$~\cite{Silverstein:2008sg}, $p=2$~\cite{Palti:2014kza,Kaloper:2008fb}, and $p=4/3, 3$~\cite{McAllister:2014mpa}, and 
$M$ depends on the number of hidden gauge sectors which non-perturbatively 
generate the potential of the axion inflaton.

We proceed to demonstrate the small-field axion inflation by the small axion 
decay constants $f_i$ in the same way as in the previous section. 
To obtain the sufficiently large number of {\it e}-folding, 
the first derivative of potential in Eq.(\ref{eq:mono_inf}),
\begin{equation}
V_{\phi} = A_1 p \phi^{p-1} -\sum_{i}\frac{A_i}{f_i} \sin\left(\frac{\phi}{f_i}+\theta_i\right),
\label{eq:first_mono_inf}
\end{equation}
is required to be smaller than the potential energy, that is, $|V_{\phi}| \ll |V|$. 
It can be realized with the region satisfying 
\begin{align}
&\phi \sim {\cal O}(1),\nonumber\\
&\sin(\phi/f_{i}+\theta_i )\sim \cos(\phi/f_{i}+\theta_i ) \sim {\cal O}(1),
\label{eq:assum_mono_1}
\end{align} 
with proper signs of $\sin(\phi/f_{i}+\theta_i )$ and $\cos(\phi/f_{i}+\theta_i )$, 
and the correlated parameters in the scalar potential,
\begin{equation}
A_1 p \sim \frac{A_i}{f_i} \sim \frac{A_j}{f_j},
\label{eq:assum_mono_2}
\end{equation}
for any $i,j=2,3,\cdots,M$.

Since the slow-roll inflation is realized under $|V_{\phi}| \simeq 0$ and $|V_{\phi \phi}| \simeq 0$ 
during the inflation, the second derivative of the potential can be estimated by employing 
the inflaton variation $\Delta \phi$, 
\begin{equation}
V_{\phi \phi} \sim V_{\phi \phi \phi} \Delta \phi \sim \left(-\sum_{i}\frac{A_i}{f_i^3} \sin\left(\frac{\phi}{f_i}+\theta_i\right)
\right)\Delta \phi,
\end{equation}
for small $f_i$. Here, we assume that all $f_i^{-3}$ in the third derivative $V_{\phi\phi\phi}$ dominate 
the second derivative $V_{\phi\phi}$. 
With the help of Eqs.~(\ref{eq:assum_mono_1}) and (\ref{eq:assum_mono_2}), the slow-roll 
parameter $\eta$ is obtained as
\begin{eqnarray}
\eta \sim  \frac{\sum_{i}\frac{A_i}{f_i^3}}{A_1} \Delta \phi  
\sim p \left(\sum_i \frac{1}{f_i^2}\right)\Delta \phi,
\end{eqnarray}
where the scalar potential during the inflation is approximately given by $V_{\rm inf} \sim A_1$ 
due to the conditions~(\ref{eq:assum_mono_1}) and~(\ref{eq:assum_mono_2}). 
Since we concentrate on the parameter space leading to the small-field inflation, 
the slow-roll parameter $\epsilon$ is 
expected to be much smaller than unity. 
It is confirmed later by checking the value of the tensor-to-scalar ratio $r = 16 \epsilon$. 
Thus, the slow-roll parameter $|\eta|$ is chosen as $10^{-2}$ to reproduce 
the observed spectral index $n_s\simeq 0.96$ reported by Planck.

By fixing $|\eta| \simeq 10^{-2}$, the tensor-to-scalar ratio is estimated by using the 
Lyth bound~Eq.(\ref{eq:Lyth_bound}), 
\begin{eqnarray}
r \sim  10^{-2} \times (\Delta \phi)^2 
\sim  10^{-6} \times \frac{1}{p^2} \left(\sum_i \frac{1}{f_i^2}\right)^{-2}
\times \left( \frac{\eta}{0.01}\right)^2,
\label{eq:r_mono}
\end{eqnarray}
from which the power of decay constants $f$ in $r$ becomes small in comparison with that in 
the multi-natural inflation.

Furthermore, we can estimate the energy scale of scalar potential during the inflation $V_{\rm \inf}$ as functions 
of axion decay constants from Eqs.~(\ref{eq:obs}) and~(\ref{eq:r_mono}),
\begin{eqnarray}
V_{\rm \inf}^{1/4} \sim  4 \times 10^{-4} \times p^{-1/2}
\left(\sum_i \frac{1}{f_i^2}\right)^{-1/2},
\label{eq:Vinf_mono}
\end{eqnarray}
and consequently the Hubble parameter $H_{\rm inf} = (V_{\rm inf}/3)^{1/2}$ becomes
\begin{eqnarray}
H_{\rm inf} \sim 10^{-7} \times p^{-1}\left(\sum_i \frac{1}{f_i^2}\right)^{-1}.
\label{eq:H_mono}
\end{eqnarray}

Finally, we estimate the inflaton mass $m_{\phi}^2$ as a function of the decay constant.
For small $f_i$, the dominant term of the second derivative, $V_{\phi \phi}$, at $\phi=0$
is evaluated by using Eq.~(\ref{eq:assum1}), $V_{\phi \phi} \sim  \sum_i \frac{A_i}{f_i^2}$ 
and hereafter the inflaton mass is estimated as 
\begin{align}
m_{\phi}^2 &=V_{\phi\phi} \sim \sum_i \frac{A_i}{f_i^2} 
\sim \frac{\sum_i\frac{A_i}{f_i^2}}{A_1}V_{\rm inf}
\sim p\left(\sum_i\frac{1}{f_i}\right) V_{\rm inf}
\sim 3\times 10^{-14} \times p^{-1}\left(\sum_i\frac{1}{f_i}\right) 
\left(\sum_i \frac{1}{f_i^2}\right)^{-2}.
\label{eq:m_mono}
\end{align}
\ \\
Let us summarize the result for two nonvanishing sinusoidal functions in Eq.~(\ref{eq:mono_inf}), 
for simplicity. 
For $f_{1} \sim f_{2} \sim f \ll 1$, the obtained physical quantities have the following decay constant 
dependence:
\begin{eqnarray}
r \sim  10^{-6} \times p^{-2} \times f^4\ , \ \ \ \ 
V_{\rm \inf}^{1/4} \sim  4 \times 10^{-4} \times p^{-1/2} \times f, \nonumber \\
H_{\rm inf} \sim 10^{-7} \times p^{-1} \times f^2\ , \ \ \ \ \ 
m_{\phi}^2 \sim 3 \times 10^{-14} \times p^{-1} \times f^3.
\label{eq:typical parameter values of type 2}
\end{eqnarray}

\begin{figure}[htbp]
  \begin{center}
    \begin{tabular}{c}

      \begin{minipage}{0.5\hsize}
        \begin{center}
          \includegraphics[clip, width=7.0cm]{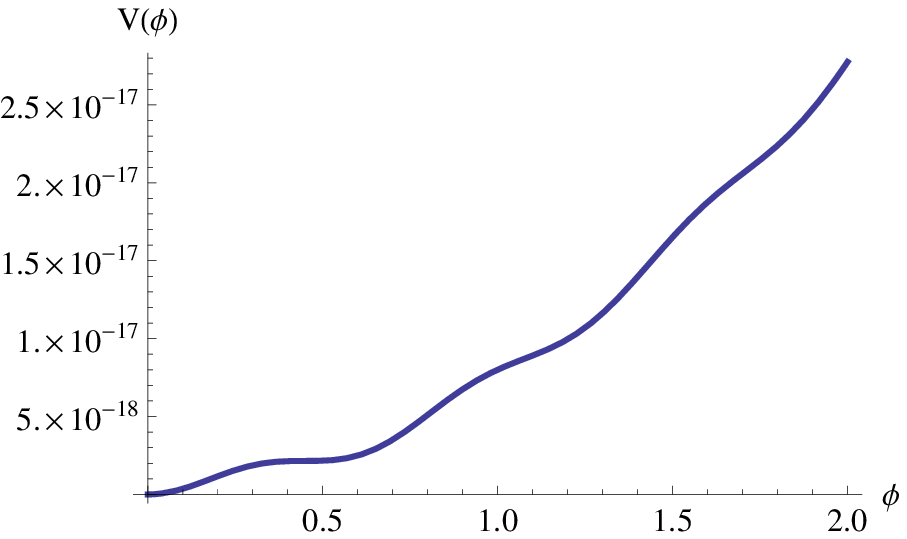}
          \hspace{1.6cm} 
        \end{center}
      \end{minipage}

      \begin{minipage}{0.5\hsize}
        \begin{center}
          \includegraphics[clip, width=7.0cm]{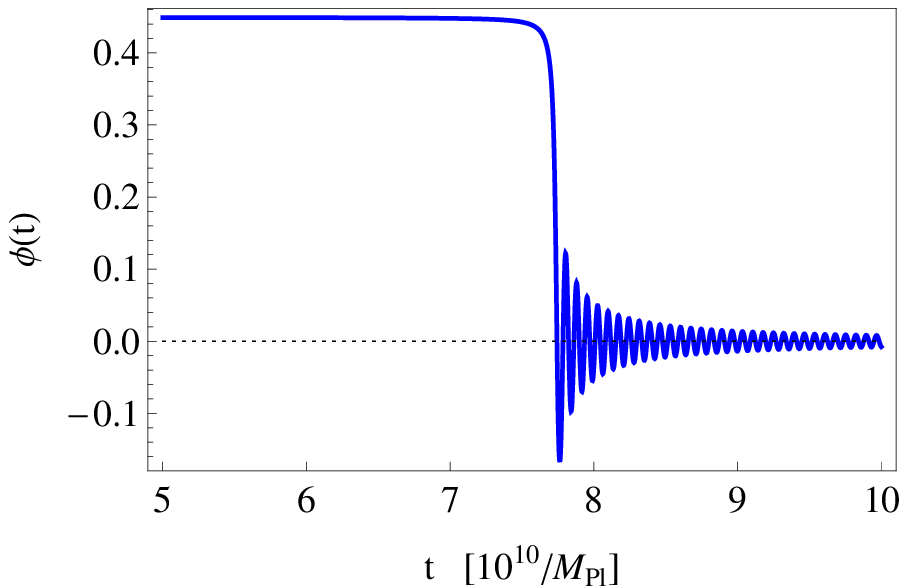}
          \hspace{1.6cm} 
        \end{center}
      \end{minipage}

    \end{tabular}
    \caption{In the left panel, the inflaton potential is drawn by setting the 
    parameters as $A_1/A_2 = 10.86169045$ and $A_2=6.30\times 10^{-19}$, whereas the right panel 
    shows the trajectory of the inflaton as a function of cosmic time $t$ for 
    the initial value of the inflaton, $\phi(0)=0.4492824$ at $t=0$.}
    \label{fig:2}
  \end{center}
\end{figure}

Following this line of thought, we show the numerical analysis for specific axion potentials.
For the illustrative purposes, we consider the axion potential with single sinusoidal function,
\begin{eqnarray}
V(\phi) &=& A_1\phi^p + A_2 \left( 1 - \cos\left(\frac{\phi}{f}\right) \right),
\label{eq:mono_simplify}
\end{eqnarray}
with $f=f_2$, $\theta_2=-\pi$ and $V_0\simeq A_2$ in Eq.~(\ref{eq:mono_inf}), 
and demonstrate the small-field axion inflation for the case with $p=2$ and 
small axion decay constants $f=0.1$. 
Fig.~\ref{fig:2} shows the inflaton potential and the trajectory of the inflaton as a function of cosmic time $t$, 
where the parameters are set as $A_1/A_2 =10.86169045$ and $A_2=6.30\times 10^{-19}$. 
By solving the equation of motion for the inflaton field, we numerically obtain the 
cosmological observables as shown in Tab.~\ref{tab:2}. 
It is then found that the analytical forms of physical quantities derived in Eqs.~(\ref{eq:r_mono})-(\ref{eq:m_mono}),
\begin{eqnarray}
r &\sim&  10^{-6} \times \frac{1}{p^2} \times f^4 \sim 3 \times 10^{-11}, \nonumber \\
V_{\rm \inf}^{1/4} &\sim&  4 \times 10^{-4} \times \frac{1}{\sqrt{p}} \times f \sim 3 \times 10^{-5}, \nonumber \\
H_{\rm inf} &\sim&  10^{-7} \times \frac{1}{p} \times f^2 \sim 5 \times 10^{-10}, \nonumber \\
m_{\phi}^2 &\sim& 3 \times 10^{-14} \times \frac{1}{p} \times f^3
\sim 8 \times 10^{-17},
\end{eqnarray}
are consistent with our obtained numerical results in Tab.~\ref{tab:2}.

\begin{table}[htb] 
 \begin{center}
  \begin{tabular}{|c|c|c|c|c|c|c|} \hline
     $N$ & $n_s$ & $r $ & 
     $m_{\phi}^2$ & $H_{\rm inf}$ &  $V_{\rm inf}^{1/4}$ & $\frac{dn_s}{d \ln k}$ \\ \hline\hline 
   60.0 & $0.9665$ & $6.60\times 10^{-11}$
   & $7.67\times 10^{-17}$& $8.46\times 10^{-10}$&  $3.83\times 10^{-5}$  & $-2.52 \times 10^{-3}$ \\ 
  50.0 & $0.9665$ & $1.55\times 10^{-10}$ 
  & $1.81 \times 10^{-16}$ & 1.30$\times 10^{-9}$ & $4.74\times 10^{-5}$ & $-1.64 \times 10^{-3}$ \\ \hline
  \end{tabular}
      \caption{The cosmological observables such as spectral index $n_s$, its running $dn_s/d\ln k$, 
      tensor-to-scalar ratio $r$, 
      Hubble scale $H_{\rm inf}$, scalar potential $V_{\rm inf}^{1/4}$ at the pivot scale and 
      the inflaton mass $m_{\phi}^2$ at the vacuum. 
      The parameters are set as $A_1/A_2 = 10.86169045$ and $A_2=6.30\times 10^{-19}$ 
      for the {\it e}-folding number $N=60$, whereas those are set as $A_1/A_2 =10.86169628$ and $A_2=6.30\times 10^{-19}$ 
      for the {\it e}-folding number $N=50$. The initial value of inflaton field is also set 
      as $\phi_{{\rm ini}}=0.4492824$ in both cases.}
    \label{tab:2}
  \end{center}
\end{table}

Tab.~\ref{tab:2} also shows numerical results of the running, $dn_s/d \ln k$, 
and these values are large and negative, again.
This value can be estimated in a way similar to the discussion in the previous section.
In this model, we can estimate $\xi ={\cal O}(10^{-3})$ again, which is independent of 
decay constants.
Thus, it is found that $dn_s/d \ln k = {\cal O}(10^{-3})$ and $d^2n_s/d \ln k^2 = {\cal O}(10^{-5})$.

Similarly, the potential (\ref{eq:mono_simplify}) with other values of $p$ and $f$ 
leads to results consistent with Eq.~(\ref{eq:typical parameter values of type 2}).

\section{Reheating temperature and dark matter abundance}
\label{sec:3}
In this section, we discuss the reheating process after the inflation dynamics. 
From now on, we assume that the inflaton axion discussed in the previous section 
couples to the gauge bosons in the standard model through tree or one-loop corrected 
gauge kinetic functions. In type IIB superstring theory on a toroidal background, it is known that 
the K\"ahler axion corresponding to the Kalb-Ramond field couples to the gauge boson at the tree-level, 
whereas the axion associated with the complex structure modulus appears in the 
gauge kinetic function at the one-loop level~\cite{Lust:2003ky,Blumenhagen:2006ci}. 
In both cases, the inflaton decays into the gauge bosons $g^{(a)}$ with $a=1,2,3$ 
corresponding to the gauge groups of 
the standard model, $U(1)_Y, SU(2)_L, SU(3)_C$, and its decay width is estimated in the 
instantaneous decay approximation, 
\begin{eqnarray}
\Gamma_{\phi} &=& \sum_{a=1}^{3} \Gamma (\phi \rightarrow 2g^{(a)}) \nonumber \\
 &\simeq& 5.8 \times 10^{-5} c^2 \left( \frac{m_{\phi}}{10^{13}\ {\rm GeV}}\right)^3 \ {\rm GeV},
\end{eqnarray}
where $c$ becomes $16\pi^2$ and unity for the K\"ahler moduli and complex structure moduli. 
When such a decay into the gauge bosons is the dominant process, 
the reheating temperature is yielded as 
\begin{eqnarray}
T_{\rm ref} = \left( \frac{\pi^2 g_\ast}{90}\right)^{-1/4}\sqrt{\Gamma_{\phi}M_{\rm Pl}} 
\simeq 6.4 \times 10^{6} c \left( \frac{m_{\phi}}{10^{13}\ {\rm GeV}}\right)^{3/2} \ {\rm GeV},
\end{eqnarray}
with the effective degrees of freedom $g_\ast=106.75$. 

\begin{figure}[htbp]
  \begin{center}
    \begin{tabular}{c}

      \begin{minipage}{0.5\hsize}
        \begin{center}
          \includegraphics[clip, width=6.5cm]{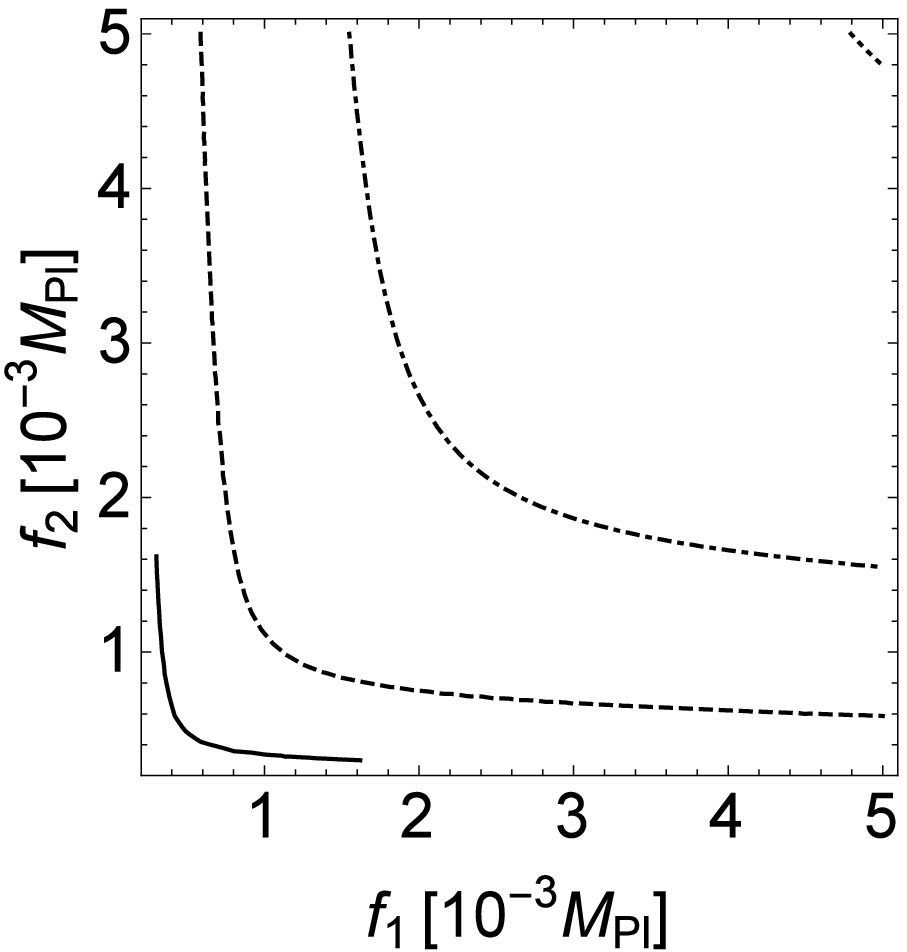}
          \hspace{1.6cm} 
        \end{center}
      \end{minipage}

      \begin{minipage}{0.5\hsize}
        \begin{center}
          \includegraphics[clip, width=6.5cm]{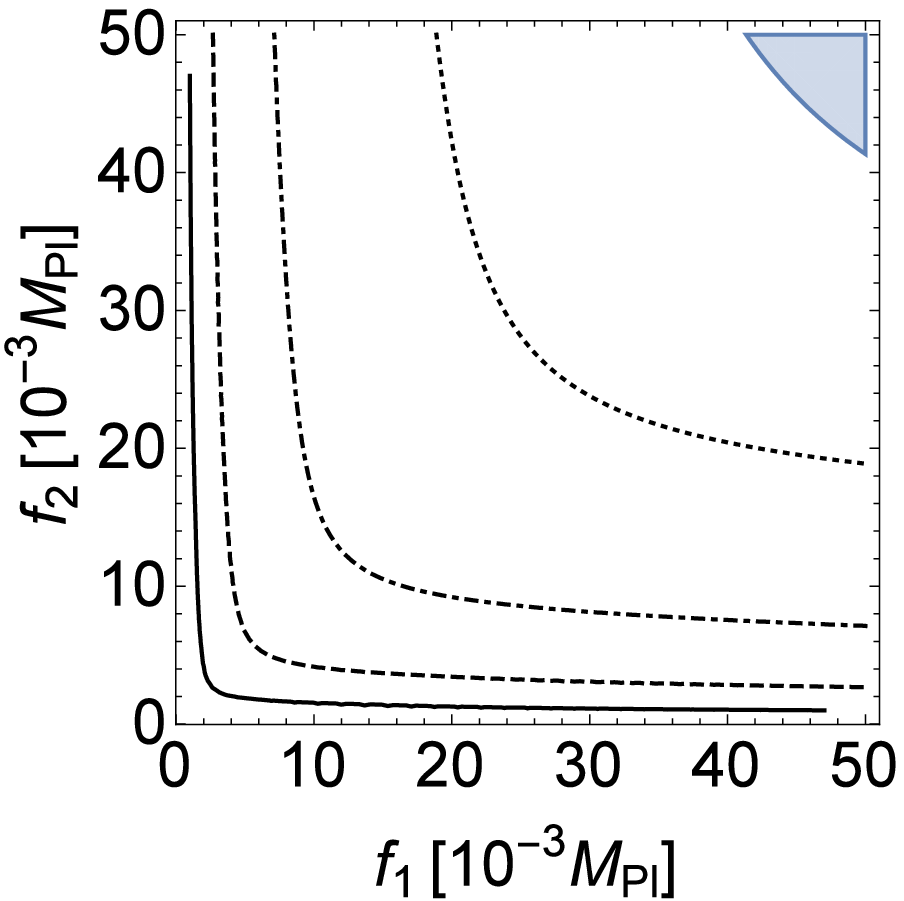}
          \hspace{1.6cm} 
        \end{center}
      \end{minipage}

    \end{tabular}
    \caption{The reheating temperature $T_{\rm reh}$ as functions of 
    two decay constants $f_{1,2}$ for the case of the axion of K\"ahler moduli 
    in the left panel and that of the complex structure modulus in the right panel. 
    In both panels, the solid, dashed, dotdashed and dotted curves 
    represent the reheating temperatures, $T_{\rm reh}=1, 10, 10^2, 10^{3}\,{\rm MeV}$, 
    respectively. The blue shaded region is excluded by the isocurvature 
    perturbation originating from the QCD axion.}
    \label{fig:3}
  \end{center}
\end{figure}

From the results in Sec.~\ref{subsec:multi}, 
the reheating temperate is yielded by the axion decay constants,
\begin{align}
T_{\rm ref} &\simeq 
5.4 \times 10^{4} c 
\left(\frac{\sum_{m=1}^M \frac{1}{f_m}}{\sum_{n=1}^M f_n}\right)^{3/4}
\left(\frac{\sum_{n=1}^M f_n}{\sum_{m=1}^M\frac{1}{f_m^2}}\right)^{3/2} \ {\rm GeV},
\end{align}
which is illustrated in Fig.~\ref{fig:3} as functions of two axion decay constants 
for the simplified multi-natural inflation in Eq.~(\ref{eq:multi_simplify}). 
We now take into account the constraint from the isocurvature 
perturbation originating from the QCD axion by Eq.~(\ref{eq:iso}) 
with $f_{\rm QCD}=10^{12}\,{\rm GeV}$ and Eq.~(\ref{eq:H_multi}) with $m=1,2$, 
which corresponds to the blue shaded region in Fig.~\ref{fig:3}. 
Here and in the following analysis, 
we employ the maximal value of the QCD axion decay constant constrained by the 
upper bound of dark matter abundance, although 
it depends on the initial misalignment angle of the axion 
and dilution mechanism after the inflation~\cite{Kawasaki:2004rx,Hattori:2015xla,Akita:2016usy}. 
As can be seen in Fig.~\ref{fig:3}, the smallest axion decay constant is bounded as 
$2\times 10^{15}\,{\rm GeV}\lesssim f\lesssim 10^{17}\,{\rm GeV}$ for 
the K\"ahler axion and $10^{16}\,{\rm GeV}\lesssim f\lesssim 10^{17}\,{\rm GeV}$ for the axion of complex structure modulus, where the lower bounds are put by $T_{\rm reh}\gtrsim {\cal O}(5)$ MeV in order not to spoil the successful BBN, 
whereas the upper bounds are set by the constraint from the 
isocurvature perturbation of the QCD axion with $f_{\rm QCD}=10^{12}\,{\rm GeV}$.

\begin{figure}[htbp]
  \begin{center}
    \begin{tabular}{c}

      \begin{minipage}{0.5\hsize}
        \begin{center}
          \includegraphics[clip, width=6.5cm]{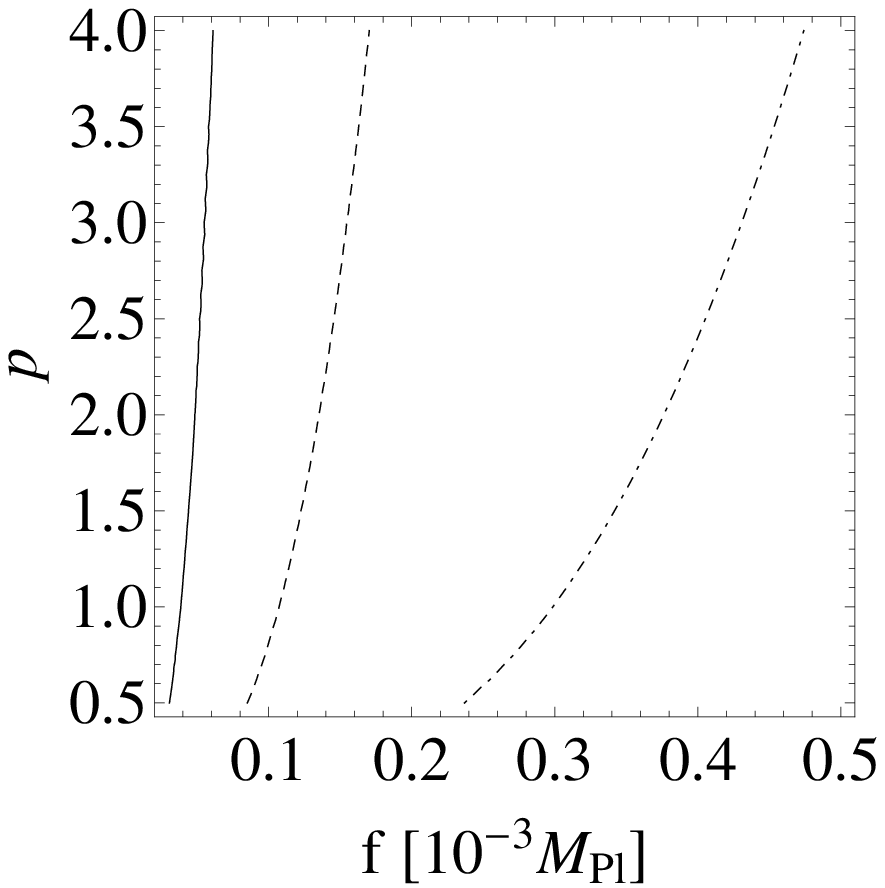}
          \hspace{1.6cm} 
        \end{center}
      \end{minipage}

      \begin{minipage}{0.5\hsize}
        \begin{center}
          \includegraphics[clip, width=6.5cm]{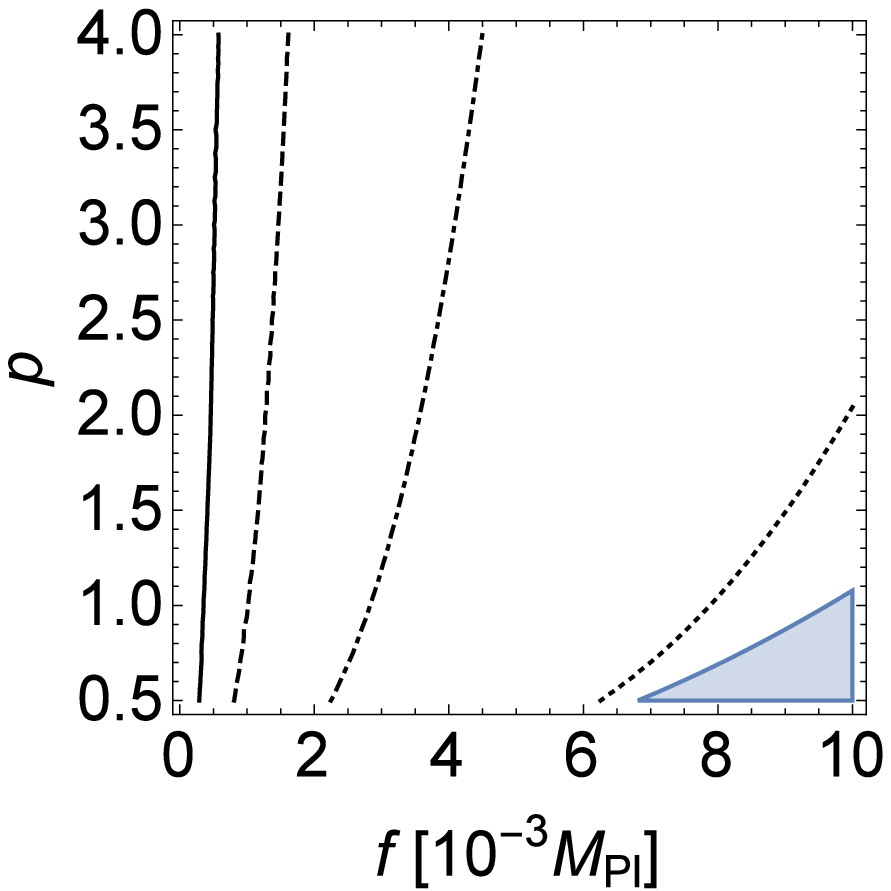}
          \hspace{1.6cm} 
        \end{center}
      \end{minipage}

    \end{tabular}
    \caption{The reheating temperature $T_{\rm reh}$ as functions of 
    decay constant $f$ and the power of he polynomial term $p$ for the case of the axion of K\"ahler moduli 
    in the left panel and that of the complex structure modulus in the right panel. 
    In both panels, the solid, dashed and dotdashed curves 
    represent the reheating temperatures, $T_{\rm reh}=1, 10, 10^2, 10^{3}\,{\rm MeV}$, 
    respectively. The blue shaded region is excluded by the isocurvature 
    perturbation originating from the QCD axion.}
    \label{fig:4}
  \end{center}
\end{figure}

Similarly, the reheating temperate for the axion monodromy inflation with 
sinusoidal functions in Sec.~\ref{subsec:mono} is also dominated by 
the axion decay constants,
\begin{align}
T_{\rm ref} &\simeq 
5.4 \times 10^{4} c \ 
p^{-3/4}\left(\sum_{i=2}^M\frac{1}{f_i}\right)^{3/4} 
\left(\sum_{i=2}^M \frac{1}{f_i^2}\right)^{-3/2}\ {\rm GeV},
\end{align}
which is illustrated in Fig.~\ref{fig:4} as functions of the axion decay constant $f=f_2\,(M=2)$ 
and the power of polynomial $p$ 
for the simplified axion monodromy inflation with sinusoidal functions in Eq.~(\ref{eq:mono_simplify}). 
In Fig.~\ref{fig:4} and in what follows, 
$p$ is considered the continuous parameter for simplicity, 
although it is a fractional number derived in a detailed string setup. 
In a similar fashion as in the multi-natural inflation, the blue shaded region in Fig.~\ref{fig:4} is excluded by the isocurvature 
perturbation originating from the QCD axion which is estimated by employing 
Eq.~(\ref{eq:iso}) with $f_{\rm QCD}=10^{12}\,{\rm GeV}$ and Eq.~(\ref{eq:H_mono}) with $f=f_2$. 
In order not to spoil the successful BBN and overproduce the isocurvature 
perturbation due to the QCD axion, 
Fig.~\ref{fig:4} gives the bounds $2\times 10^{14}\,{\rm GeV}\lesssim f\lesssim 5\times 10^{16}\,{\rm GeV}$ for 
the K\"ahler axion and $2\times 10^{15}\,{\rm GeV}\lesssim f\lesssim 5\times 10^{16}\,{\rm GeV}$ for the axion of the complex structure modulus. 
Note that the smaller $f_{\rm QCD}$ gives 
the tight upper bound on $f$ from the isocurvature perturbation of the QCD axion. 
As a result, 
these regions correspond to the typical decay constant for the closed 
string axions~\cite{Choi:1985je,Banks:2003sx,Svrcek:2006yi}. 
That is surprisingly interesting. 
Although we focus on the simplified axion potentials in Eqs.~(\ref{eq:multi_simplify}) and~(\ref{eq:mono_simplify}), 
such severe constraints for the axion decay constant are also applied to the general form 
of the axion potential. Indeed, the larger axion decay constants lead to the large 
Hubble scale given in Eqs.~(\ref{eq:H_multi}) and~(\ref{eq:H_mono}).

From these considerations, the low-scale axion decay constants realizing 
the successful small-field axion inflations generically predict the low reheating 
temperature. 
It implies that the freeze-out temperature of dark matter would be smaller 
than the reheating temperature and consequently 
the dark matter yield is determined by the non-thermal process from the inflaton 
decay
\begin{align}
\frac{n_{\rm dm}}{s}\simeq \frac{3T_{\rm reh}}{4m_{\phi}}{\rm Br}_{\rm dm}
\simeq 4.8\times 10^{-7}c\,{\rm Br}_{\rm dm}\left( \frac{m_{\phi}}{10^{13}\ {\rm GeV}}\right)^{1/2},
\end{align}
where $n_{\rm dm}$ is the number density of dark matter, $s$ 
is the entropy density of the Universe and ${\rm Br}_{\rm dm}$ 
is the branching ratio from the inflaton to dark matter. 
The relic abundance of dark mater is then given in terms of 
the ratio of the critical density to the current entropy density 
of the Universe $\rho_{\rm cr}/s_0\simeq 3.6h^2\times 10^{-9}$,
\begin{align}
\Omega_{\rm dm}h^2 \simeq m_{\rm dm}\frac{n_{\rm dm}}{s}\frac{s_0}{\rho_{\rm cr}}
\simeq 1.3c\left(\frac{m_{\rm dm}}{100\,{\rm GeV}}\right)
\left(\frac{{\rm Br}_{\rm dm}}{10^{-4}}\right)
\left( \frac{m_{\phi}}{10^{13}\ {\rm GeV}}\right)^{1/2},
\end{align}
with $h$ being the dimensionless Hubble parameter. 

\begin{figure}[htbp]
  \begin{center}
    \begin{tabular}{c}

      \begin{minipage}{0.4\hsize}
        \begin{center}
          \includegraphics[clip, width=6.5cm]{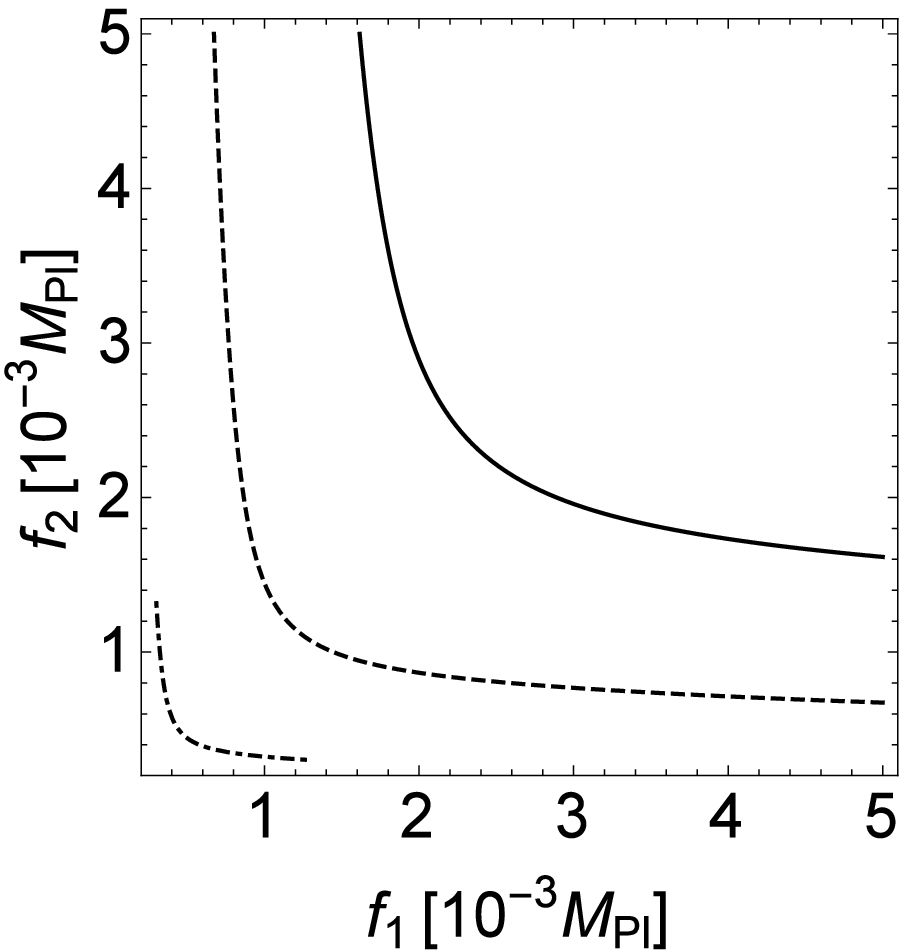}
          \hspace{1.6cm} 
        \end{center}
      \end{minipage}

      \begin{minipage}{0.4\hsize}
        \begin{center}
          \includegraphics[clip, width=6.5cm]{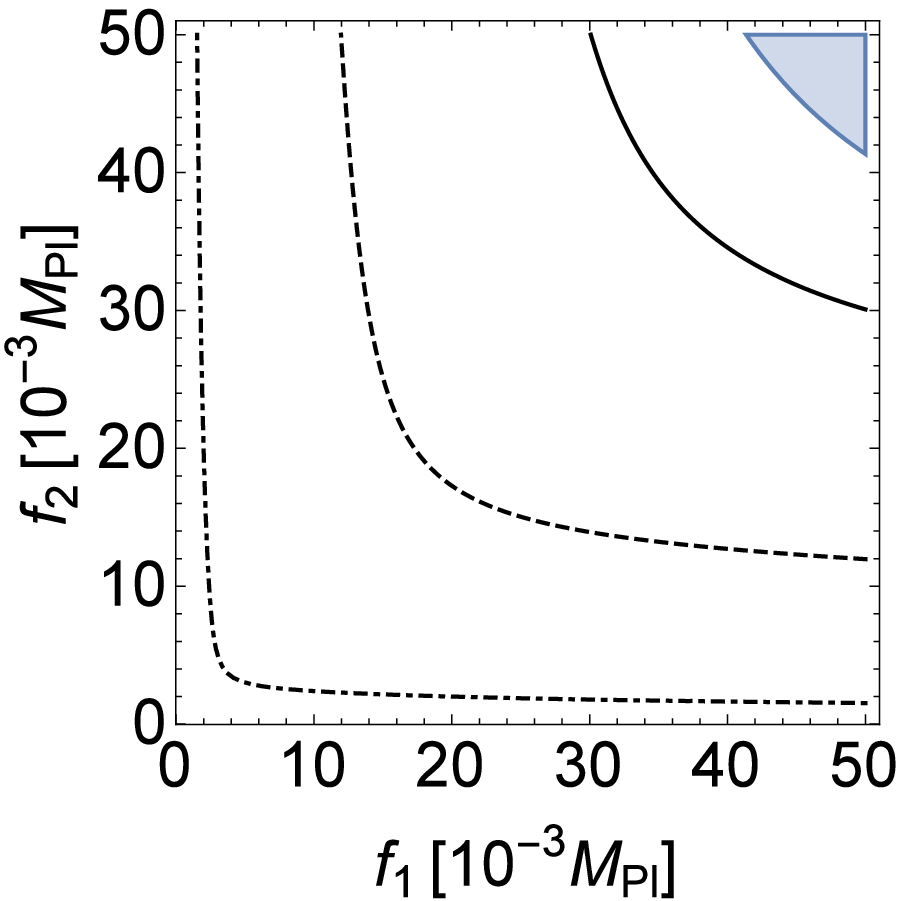}
          \hspace{1.6cm} 
        \end{center}
      \end{minipage}

    \end{tabular}
    \caption{The dark matter abundance $\Omega_{\rm dm}h^2=0.1$ as functions of 
    two decay constants $f_{1,2}$ for the case of the axion of the K\"ahler moduli 
    in the left panel and that of the complex structure modulus in the right panel. 
    In the left panel, the black solid, dashed, and dotdashed curves 
    are drawn by setting $m_{\rm dm}{\rm Br}_{\rm dm}=10^{-2}, 0.02, 0.05\,{\rm GeV}$, 
    respectively, whereas, in the right panel, the black solid, dashed and dotdashed curves 
    are drawn by setting $m_{\rm dm}{\rm Br}_{\rm dm}=10^{-1}, 0.2, 1\,{\rm GeV}$, 
    respectively. 
     The blue shaded region is excluded by the isocurvature 
    perturbation originating from the QCD axion with $f_{\rm QCD}=10^{12}\,{\rm GeV}$.}
    \label{fig:5}
  \end{center}
\end{figure}

\begin{figure}[htbp]
  \begin{center}
    \begin{tabular}{c}

      \begin{minipage}{0.4\hsize}
        \begin{center}
          \includegraphics[clip, width=6.5cm]{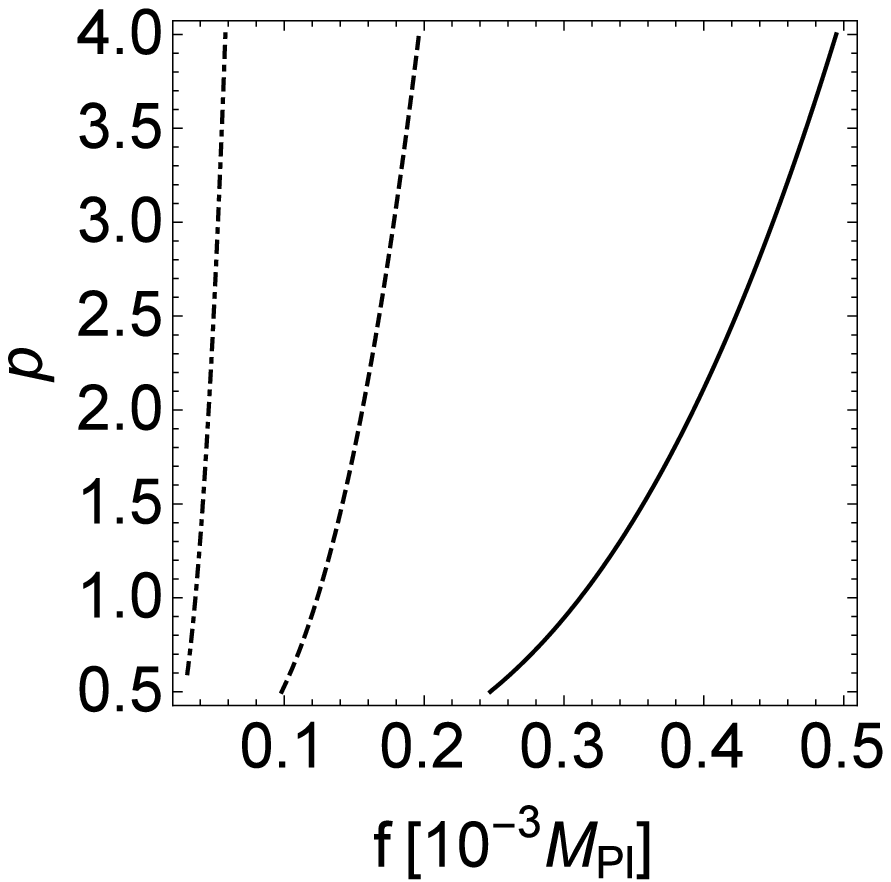}
        \end{center}
      \end{minipage}

      \begin{minipage}{0.4\hsize}
        \begin{center}
          \includegraphics[clip, width=6.5cm]{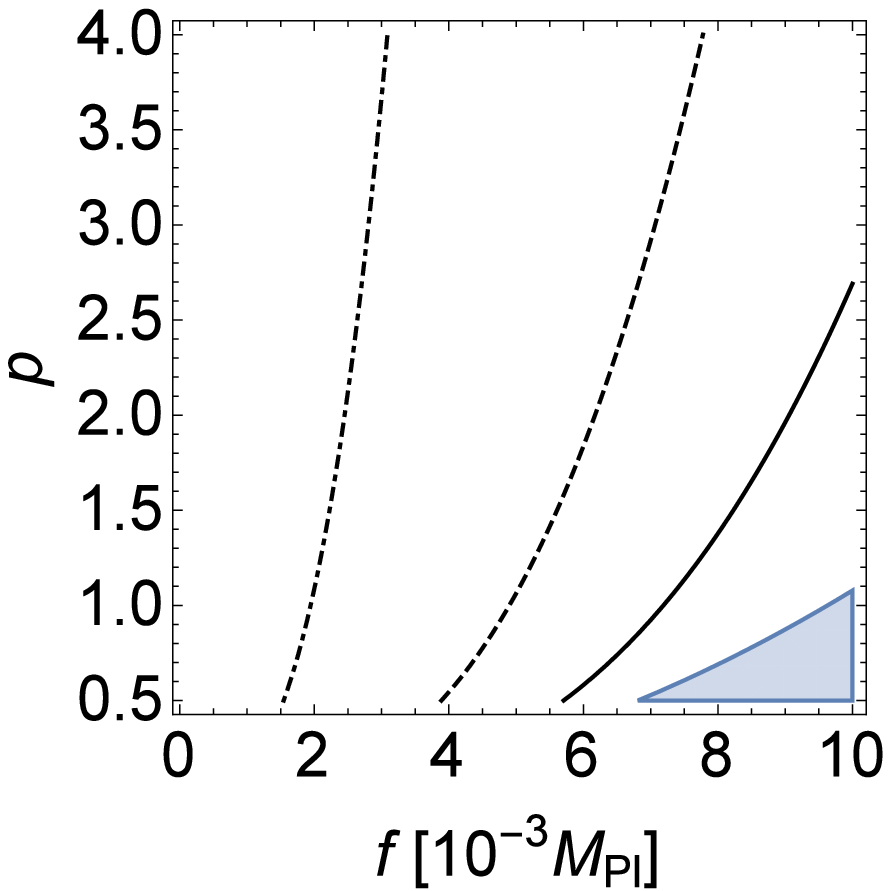}
        \end{center}
      \end{minipage}

    \end{tabular}
    \caption{The dark matter abundance $\Omega_{\rm dm}h^2=0.1$ as functions of 
    decay constant $f$ and the power of polynomial term $p$ for the case of the axion of the K\"ahler moduli 
    in the left panel and that of the complex structure modulus in the right panel. 
    In the left panel, the black solid, dashed and dotdashed curves 
    are drawn by setting $m_{\rm dm}{\rm Br}_{\rm dm}=10^{-2},0.02, 0.05\,{\rm GeV}$, 
    respectively, whereas, in the right panel, the black solid, dashed, and dotdashed curves 
    are drawn by setting $m_{\rm dm}{\rm Br}_{\rm dm}=0.15,0.2,0.4\,{\rm GeV}$, 
    respectively. 
     The blue shaded region is excluded by the isocurvature 
    perturbation originating from the QCD axion with $f_{\rm QCD}=10^{12}\,{\rm GeV}$.}
    \label{fig:6}
  \end{center}
\end{figure}

From now on, we for simplicity assume that the current dark matter abundance 
mainly consists of the QCD axion compared with another cold dark matter.
In Figs.~\ref{fig:5} and~\ref{fig:6}, we plot the dark matter abundance 
for the simplified multi-natural inflation in Eq.~(\ref{eq:multi_simplify}),
\begin{align}
\Omega_{\rm dm}h^2 
\simeq 0.27c\left(\frac{m_{\rm dm}}{100\,{\rm GeV}}\right)
\left(\frac{{\rm Br}_{\rm dm}}{10^{-4}}\right)
\left(\frac{\sum_{m=1}^2 \frac{1}{f_m}}{\sum_{n=1}^2 f_n}\right)^{1/4}
\left(\frac{\sum_{n=1}^2 f_n}{\sum_{m=1}^2\frac{1}{f_m^2}} \right)^{1/2},
\end{align}
and for the simplified axion monodromy inflation with sinusoidal functions in Eq.~(\ref{eq:mono_simplify}), 
\begin{align}
\Omega_{\rm dm}h^2 
\simeq 0.27c\left(\frac{m_{\rm dm}}{100\,{\rm GeV}}\right)
\left(\frac{{\rm Br}_{\rm dm}}{10^{-4}}\right)
2^{-1/4}f^{3/4},
\end{align}
respectively. 
The relic dark matter abundance should be less than $\Omega_{\rm dm}h^2 \simeq 0.12$ 
reported by Planck in order to not overclose our Universe~\cite{Ade:2015lrj}.
Although these predictions depend on the branching ratio ${\rm Br}_{\rm dm}$ and dark matter mass, 
$\Omega_{\rm dm}h^2 < 0.12$ in Figs.~\ref{fig:5} and~\ref{fig:6} is achieved in both inflation models. 
For example, in simplified multi-natural inflation in Eq.~(\ref{eq:multi_simplify}), 
$\Omega_{\rm dm}h^2 < 0.12$ can be realized under 
e.g., ${\rm Br}_{\rm dm}< {\cal O}(10^{-4})$ and $m_{\rm dm}\simeq 100\,{\rm GeV}$ 
with $f_{1,2}\simeq 2\times 10^{-3}\,{\rm M_{\rm Pl}}$ 
for the K\"ahler axion and ${\rm Br}_{\rm dm}< {\cal O}(10^{-3})$ and $m_{\rm dm}\simeq 100\,{\rm GeV}$ 
with $f_{1,2}\simeq 2\times 10^{-2}\,{\rm M_{\rm Pl}}$ for the axion of the complex structure modulus, 
whereas in axion monodromy inflation with sinusoidal functions in Eq.~(\ref{eq:mono_simplify}), 
$\Omega_{\rm dm}h^2 < 0.12$ can be realized under 
e.g., ${\rm Br}_{\rm dm}< {\cal O}(10^{-4})$ and $m_{\rm dm}\simeq 100\,{\rm GeV}$ 
with $f\simeq 3\times 10^{-4}\,{\rm M_{\rm Pl}}$ 
for the K\"ahler axion and ${\rm Br}_{\rm dm}< {\cal O}(10^{-3})$ and $m_{\rm dm}\simeq 100\,{\rm GeV}$ 
with $f\simeq 10^{-2}\,{\rm M_{\rm Pl}}$ for the axion of complex structure modulus. 
However, the low-scale inflation requires enough baryon asymmetry 
to reproduce the current baryon asymmetry of our Universe. 
To explain the relic baryon asymmetry, we could combine our inflation models with the baryogenesis scenario, 
e.g., the Affleck-Dine mechanism~\cite{Affleck:1984fy,Dine:1995kz}.
It would be studied in a future work.


\clearpage
\section{Conclusion}
\label{sec:con}
We have discussed the general class of small-field axion inflation, which is 
the mixture of polynomial and sinusoidal functions with an emphasis on the 
small axion decay constant compared with the Planck scale. 
In contrast to the large-field axion inflation such as the natural inflation~\cite{Freese:1990rb} 
and axion monodromy inflation~\cite{Silverstein:2008sg}, 
the small-field axion inflation predicts 
the small amount of primordial gravitational waves and low inflation scale. 
This class of inflation models is motivated by the weak gravity conjecture, 
which prohibits the trans-Planckian axion decay constant and 
the constraint from isocurvature perturbation due to the QCD axion. 
When the axion decay constants and parameters in the scalar potential 
satisfy the certain conditions leading to the successful small-field axion inflations 
as discussed in Sec.~\ref{sec:2}, 
we find that the cosmological observables are written in terms of the axion decay constants 
in a systematic way. 

Furthermore, the axion decay constant is severely constrained within the range 
$10^{14}\,{\rm GeV}\lesssim f\lesssim 10^{17}\,{\rm GeV}$, 
where the lower bounds are put by $T_{\rm reh}\gtrsim {\cal O}(5)$ MeV in order not to spoil the successful BBN, 
whereas the upper bounds are set by the constraint from the 
isocurvature perturbation due to the QCD axion with $f_{\rm QCD}=10^{12}\,{\rm GeV}$. 
This constrained axion decay constant naturally appears in 
the string theory, when our discussed axion corresponds to the closed string 
axion~\cite{Choi:1985je,Banks:2003sx,Svrcek:2006yi}. 
Although the parameters in the axion potential should be properly chosen 
to achieve a flat enough direction in the axion potential, 
the small-field axion inflation is attractive from the 
theoretical and phenomenological points of view.

\section*{Acknowledgements}

 T.K. is supported in part by
the Grant-in-Aid for Scientific Research No.~26247042 from the Ministry of Education,
Culture, Sports, Science and Technology  in Japan.


\end{document}